# Dick effect in a pulsed atomic clock using coherent population trapping

Jean-Marie Danet, Michel Lours, Stéphane Guérandel, Emeric de Clercq

LNE-SYRTE, Observatoire de Paris, CNRS, UPMC, 61 Avenue de l'Observatoire, 75014 Paris, France.

Abstract. The Dick effect can be a limitation of the achievable frequency stability of a passive atomic frequency standard when the ancillary frequency source is only periodically sampled. Here we analyze the Dick effect for a pulsed vapor cell clock using coherent population trapping (CPT). Due to its specific interrogation process without atomic preparation nor detection outside of the Ramsey pulses, it exhibits an original shape of the sensitivity function to phase noise of the oscillator. Numerical calculations using a three-level atom model are successfully compared with measurements; an approximate formula of the sensitivity function is given as an easy-to-use tool. A comparison of our CPT clock sensitivity to phase noise with a clock of the same duty cycle using a two-level system reveals a higher sensitivity in the CPT case. The influence of a free-evolution time variation and of a detection duration lengthening on this sensitivity is studied. Finally this study permitted to choose an adapted quartz oscillator and allowed an improvement of the clock fractional frequency stability at the level of 3.2×10-13 at 1s.

## I. INTRODUCTION

Many passive frequency standards are operated sequentially. Each cycle is typically composed of three steps: atomic preparation (e.g. cooling, optical pumping, etc.), clock transition interrogation (interaction of the atoms with the EM field issued from the local oscillator (LO)), and detection of the atomic response. Thus the atoms are interrogated only during a finite time. Therefore, the atomic response and the derived correction signal applied to the LO are available at the end of each cycle. It has been shown that the lack of information on the frequency of the LO due to the sampling process can lead to a limitation of the short-term frequency stability of the atomic standard [1-5]. This effect, known as Dick effect, originates from the down-conversion of the LO intrinsic frequency noise at Fourier frequencies higher than the interrogation frequency.

A great deal of work has been done for investigating the Dick effect in standards using a Ramsey interrogation technique [1-5]. In such a standard, after a time of preparation, atoms undergo two successive microwave pulses of the same length separated by a free evolution time in the dark. The atomic response is then detected by another interaction. Here we report an investigation on the limitation of the short term stability by Dick effect in a clock using coherent population trapping (CPT) and a Ramsey interrogation. In a CPT Ramsey clock the sequence of one cycle is very different of the usual Ramsey sequence. There is no atomic preparation neither detection outside the Ramsey pulses. The first light pulse of the Ramsey interrogation pumps the atoms in a dark state. After a free evolution time a second very short pulse produces the Ramsey fringes and is used for detection in the same time by monitoring the fluorescence light or the transmitted power.

The specificity of CPT Ramsey fringes will be addressed in the next Section, with description of the experimental set-up. Section III will present the numerical calculation of the sensitivity function. Its comparison with the one calculated for a two-level system equivalent clock is made and the consequences of the shape differences are discussed. Section IV will firstly deal with the experimental confirmation of the mathematical model. Secondly the possibility of a time sequence optimisation is discussed through the study of the Dick effect dependence with two parameters, the free evolution time and the detection duration. Finally, the evaluation of Dick effect in our setup, the improvement on the microwave noise and on the stability that have followed will be presented in section V and VI.

## II. RAMSEY CPT CLOCK

The CPT phenomenon occurs when two phase coherent laser beams connect two atomic ground states to a common excited state forming a so-called $\Lambda$ scheme [6, 7]. On resonance, when the laser frequency difference is equal to the ground state splitting, the atoms are pumped in a trapping state: a coherent state superposition of both ground states which does not absorb the light, called dark state. The light absorption and the fluorescence decrease or vanish. The ultimate resonance linewidth is fixed by the lifetime of the ground state coherence. Narrow microwave resonances can thus be observed without the need of a resonant cavity. Atomic clocks can be based on CPT in alkali-metal atoms [8-11]. The two levels of the clock transition are connected to a common excited level of the D1 or D2 line by two laser frequencies. Narrower linewidths can be obtained by combining CPT and a Ramsey interrogation technique [12, 13]. In this case a first laser pulse pumps the atoms in the dark state, which is a steady-state. It is analog to optical pumping, using spontaneous emission process. There is no equivalent of a π/2 pulse like in traditional microwave Cs clocks, atomic beam or fountains [14, 15]. The second Ramsey pulse must be very short, otherwise the atoms are pumped again in the steady state and the fringes will vanish [13, 16, 17].

The same pulse is also used for signal detection, by means of fluorescence or absorption. A cycle of operation of a Ramsey CPT clock is then very simple and different from that of traditional clocks. It consists of a long pulse and a short pulse separated by a delay time, see Fig. 1.

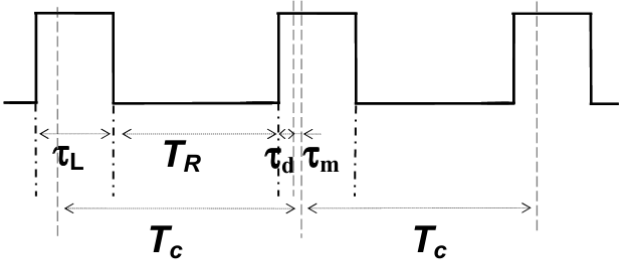

Fig. 1. Time sequence of a CPT Ramsey clock (not to scale). $\tau_L$ is the length of each laser pulse, $T_R$ is the free evolution time. The signal is detected after a time $\tau_d$ at the beginning of each pulse, and averaged during a time $\tau_m$. The length of a cycle is $T_c = \tau_L-(\tau_d+\tau_m)+T_R+\tau_d+\tau_m = \tau_L+T_R$.

In clock operation the Cs vapor is illuminated by a train of laser pulses of length $\tau_L$ (typically 2 ms in our set-up), separated by a Ramsey time $T_R$ (4 ms) during which the atoms evolve freely in the dark. At the beginning of each pulse the signal is detected after a time $\tau_d$ (100 µs), and averaged during a time $\tau_m$ (25 µs), corresponding to the end of each cycle. The atoms are then pumped again in the dark state. The full length of a cycle is $T_c = \tau_L + T_R$. The resulting Ramsey fringes are shown in Fig 2, for different Ramsey time, 1, 4 and 8 ms.

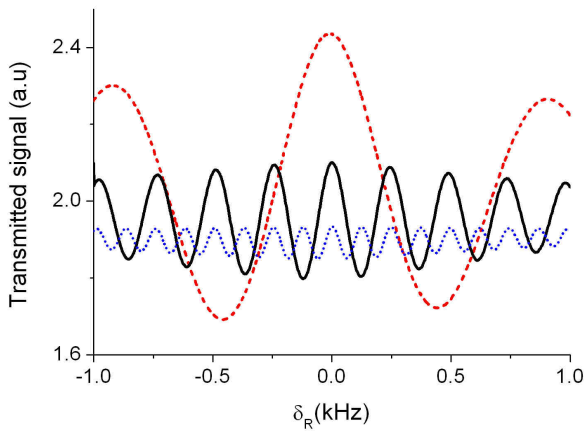

Fig. 2. Experimental Ramsey fringes for different free-evolution times: 1, 4 and 8 ms for dashed, solid and dotted lines, respectively.

A schematic view of the pulsed CPT Cs clock developed in our laboratory is shown in Fig. 3. The two frequencies are generated by two phase-locked extended cavity diode lasers (ECDL). The master laser is frequency locked to the (F = 4 - F' = 4) hyperfine component of the Cs D1 line by saturated absorption in an auxiliary vapor cell. The slave laser is phase-locked to the master laser with a frequency offset tunable around 9.192 GHz. For this purpose, the two beams are superimposed and detected by the fast photodiode PD0, the 9.192 GHz beat note is compared in phase to a 9.392 GHz signal issued from a low noise frequency synthesizer. The resulting 200 MHz signal is then used in the phase lock loop (PLL) to lock the slave laser.

The frequency synthesizer [18] is build as follows. A 100-MHz local oscillator (LO) is phase locked on the 100-MHz signal generated by an H maser of the laboratory. The LO signal is frequency multiplied to produce a 9 400 MHz signal, which is mixed with a 7.4 MHz signal to yield the tunable 9 392 MHz signal. The 7.4 MHz signal is produced by a home-made direct digital synthesizer (DDS), whose phase and frequency can be controlled by a computer (PC).

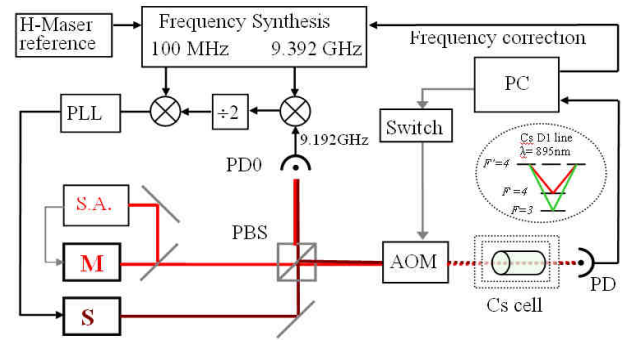

Fig. 3. Scheme of the experimental set-up. The master laser (M) is frequency locked on Cs resonance by a saturated absorption scheme (SA). The slave laser (S) is phase-locked on the master laser. PBS polarizing beam splitter, PD0 fast photodiode, PLL phase-lock loop. The laser beams are switched by the acousto-optic modulator (AOM) driven by the computer (PC). The transmitted power is recorded by the photodiode PD. The inset shows the involved Cs energy levels.

The two superimposed laser beams are switched on and off by an acousto-optic modulator (AOM) before being expanded to a 1.7 cm diameter and travelling across the Cs cell. The beams of 5 µW/mm² each are linearly and orthogonally polarized. The Cs cell, 5 cm long and 2 cm in diameter, is filled with a buffer gas at the pressure of 21 torr (Ar-N2 mixture at a pressure ratio 0.6 [19]). The cell is temperature controlled around 29°C within the mK level. It is surrounded by a solenoid applying a static magnetic field, and two magnetic shields. The transmitted power is recorded by a low-noise Si photodiode (PD). The resulting digitalized signal is processed by a computer (PC) which drives the 7.4 MHz synthesizer as well as the AOM switch.

To lock the synthesizer frequency, the probe frequency (PF) is held at resonance, and the phase is square wave modulated by ± π/2 during the dark period. The error signal is computed by the PC which can lock the LO frequency or equivalently the synthesizer frequency. Usually the LO is phase-locked on the maser and the synthesizer frequency is locked on the resonance. Its frequency is proportional to the frequency difference between the maser and the Cs clock and the Allan variance is computed from its record.

## III. SENSITIVITY FUNCTION

It has been shown that the lowest limit to the achievable stability of a sequentially operated frequency standard, induced by the interrogation frequency noise, is given by [2-5]:

$$\sigma_y^2(\tau) = \frac{1}{\tau} \sum_{m=1}^{\infty} \left[ \frac{g_m^2}{g_0^2} S_y^f\left(\frac{m}{T_c}\right) \right], \quad (1)$$

where $\sigma_y^2(\tau)$ is the fractional Allan variance for an averaging time $\tau$, $S_y^f(m/T_c)$ is the one-sided power spectral density (PSD) of the relative frequency fluctuations of the free running probe frequency at Fourier frequency $m/T_c$, i.e. the harmonic frequencies of the clock interrogation frequency $f_i$. The parameters $g_m$ and $g_0$ are defined from the sensitivity function $g$ as follows:

$$g_0 = \frac{1}{T_c} \int_0^{T_c} g(\theta) d\theta$$
$$g_m^2 = (g_m^s)^2 + (g_m^c)^2,$$
$$\begin{pmatrix} g_m^s \\ g_m^c \end{pmatrix} = \frac{1}{T_c} \int_0^{T_c} \begin{pmatrix} \sin(2\pi m \theta / T_c) \\ \cos(2\pi m \theta / T_c) \end{pmatrix} g(\theta) d\theta \quad (2)$$

### A. Sensitivity Function for CPT Systems

The sensitivity function $g(t)$ is the sensitivity of the response of the atomic system to an infinitesimal phase step of the interrogation oscillator:

$$g(t) = \lim_{\Delta\varphi \to 0} \delta S(t, \Delta\varphi) / \Delta\varphi, \quad (3)$$

where $\delta S(t, \Delta\varphi)$ is the change of the signal $S$ induced by a phase step $\Delta\varphi$ arising at time $t$. The stability limit is then entirely determined by the harmonic content of the PF PSD and of the sensitivity function which characterizes the atomic response during an interrogation cycle.

In clocks using a transition in a "two-level atom" driven by an electromagnetic (EM) field, $S(t, \Delta\varphi)$ and the sensitivity function can be calculated analytically. For the CPT phenomenon which involves three atomic levels and two EM fields, see inset in Fig. 3, there is no known analytical solution available. The sensitivity function must be computed numerically from the evolution equation of the density matrix $\rho$ [20, 6]. A complete calculation should take into account all the involved Zeeman sublevels, however good results are obtained with a simplified three-level system [10]. We consider the two clock levels of the Cs ground state, and a single excited level. The related optical Bloch equations are given in the Appendix. In order to compute $\delta S(t, \Delta\varphi)$, (A2) is numerically integrated over the first pulse duration, the free evolution time, and the second pulse duration. A phase step $\Delta\varphi$ between the phases of the two laser fields is introduced at a variable time $t$.
The sensitivity function computed with the following parameters values: first pulse 2 ms, free evolution time 4 ms, second pulse 25 µs after 100 µs delay, is shown in Fig. 4.

In the case where $\tau_d$ and $\tau_m \ll T_R$, $g(t)$ is well approximated by neglecting the pre-detection bump and by approximating the sharp exponential decrease that occurs during the detection average by a linear function:

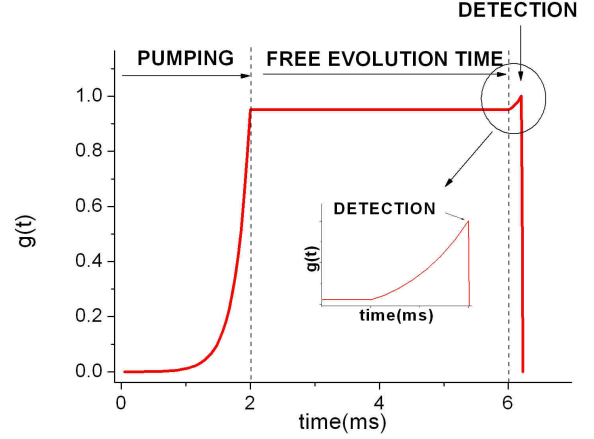

Fig. 4. Calculated sensitivity function of a pulsed CPT system. $g$ is normalized to 1 at its maximum. The inset is a zoom of the second pulse part.

$$g(t) = \begin{cases} \exp((t-\tau_L)(\gamma_C + 1/\tau_P)) & 0 \le t \le \tau_L \\ 1 & \tau_L \le t \le \tau_L + T_R + \tau_d \\ 1 - \dfrac{t-(\tau_L + T_R + \tau_d)}{\tau_m} & \tau_L + T_R + \tau_d \le t \le T_C \end{cases}, \quad (4)$$

where $\tau_P = \Gamma/\Omega^2$ is a typical pumping time, with $\Gamma$ the relaxation rate of the population of the excited level, $\Omega^2$ the quadratic sum of the Rabi frequencies related to both optical transitions, and $\gamma_c$ is the coherence relaxation rate. This approximation corresponds to our working conditions and allows the estimation of the Allan standard deviation limited by Dick effect within 1% of the non-approximated value.

In Fig. 4, it can be observed that $g(t)$ doesn't reach its maximum value during the pumping process. Thus, during the comparison of the atomic frequency and the LO frequency, this pumping duration is equivalent to a dead time. Therefore, the LO noise will not be detected by the atoms and thus the LO frequency will not be corrected. The dead time duration is compared to phase noise sensitive duration by calculating the duty cycle: $T_R/T_c$. In our case its value is 0.66. The Dick effect would vanish for a duty cycle of 1.

### B. Comparison Between CPT and a Classical Two-Level System

A two-level setup similar to our three-level one in term of duty cycle and performance goal is the pulsed Rb clock with optical pumping [21], for which the optical pumping occurs during 400 µs and the interrogating sequence is composed by two microwave π/2 pulses of 0.4 ms duration each, separated by a 3.3 ms free evolution time. A detection duration of 150 µs then follows. In Fig.

5 we compare the Fourier coefficients of the sensitivity function of both clocks, with a third one where $g(t)$ would be a simple rectangular window function, see inset in Fig. 5. At low Fourier frequencies, a common -20dB/decade slope is observed. At higher frequencies, the main difference is the presence or absence not of a -40 dB/decade slope and its starting Fourier frequency. The lower it is, the better the weighting of the LO noise will be. Because of the steep slope of $g(t)$ during the detection process in the case of CPT, (see curve (b) in the inset of Fig 5), this slope appears at a $m$ value three times higher than for the two-level system; this leads to a higher sensitivity to the LO noise in the CPT case.

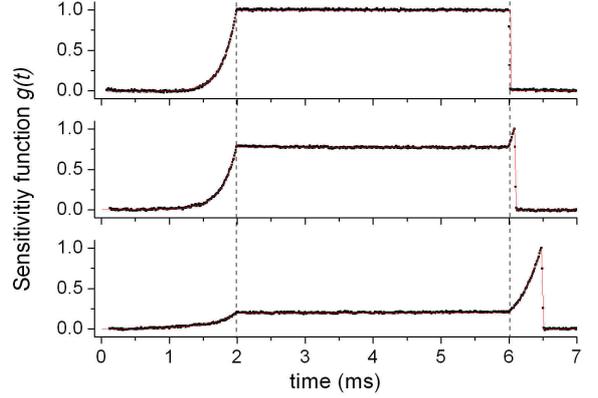

Fig. 6. Measured (dots) and calculated (solid line) pulsed CPT sensitivity function for three different $\tau_d$, from top to bottom $\tau d$ =10 µs, 80 µs, 500 µs. $g(t)$ is normalized to 1 at its maximum value.

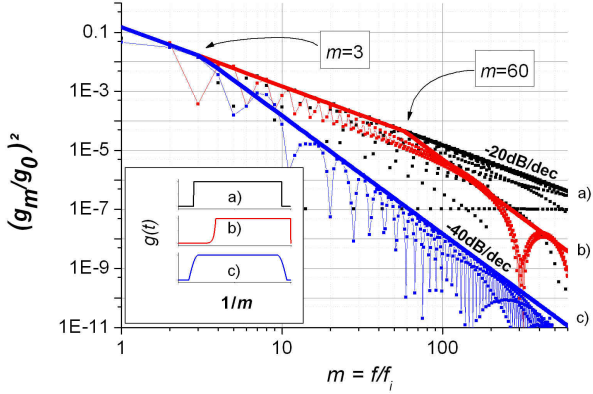

Fig. 5. Fourier coefficient of $g(t)$ for different $g(t)$ shapes: dots = results of the calculations; lines: asymptotic behaviour. (a) Square window type (upper curve, black on line), (b) CPT type (middle curve, red on line), (c) two-level type (lower curve, blue on line). Inset: $g(t)$ shapes, a, b and c refer to the same case as for the Fourier coefficients.

## IV. SENSITIVITY FUNCTION AND FREQUENCY STABILITY

### A. Mathematical Model Validation

Two main experimental tests of the calculations have been successfully realized. The first was devoted to check the $g(t)$ shape and the second to test the consistency between the calculated Dick effect contribution and our clock frequency instability measurements. The sensitivity function $g(t)$ has been measured, as in [22], by applying a 157 mrad phase step generated at different times $t$ by the computer-controlled DDS. $g(t)$ is deduced from the signal difference with and without phase step as a function of $t$.

It is noticeable that the CPT pulsed sensitivity function shows an original shape which does not reach its maximum value during the Ramsey time, unlike clocks based on a two-level atomic scheme. Indeed, because of the small delay $\tau_d$ between the turning-on of the lasers and the detection start, our setup's most sensitive instant occurs just before detection. By increasing $\tau_d$ we have been able to confirm such a behavior as shown in Fig. 6.

To validate the Dick effect contribution to the total frequency stability of our set-up, the first step consisted of injecting white phase noise in the microwave chain by adding voltage white noise on the DDS signal (see Section II) using a summing amplifier. From the new phase noise spectrums (see Fig. 7 top), we have calculated the Dick effect contribution and compared it with the measured total clock frequency instability. Both are in very good agreement as shown in Fig. 7 (bottom). The threshold point at -102 dBrad²/Hz marks the level at which the multiplication chain noise starts to dominate the microwave noise budget.

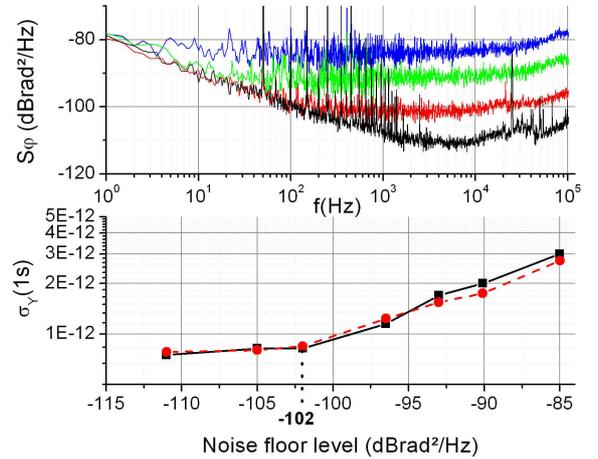

Fig. 7. (top) Phase PSD of the multiplication chain for different injected white phase noise , from top to bottom: -84, -91, -102, and less than -111 dBrad²/Hz. (bottom) Fractional Allan deviation at 1s: square and solid line = calculated values using the noised phase noise PSD; dots and dashed line = measured values. The solid lines are a guide to the eye.

### B. Minimizing Dick Effect With a Time Sequence Optimization

To provide us with a better understanding of Dick effect in a CPT based system, we have studied how the phase sensitivity could be decreased by an optimized time sequence. Two parameters are investigated, the detection windows and the duty cycle. However, the difficulty of such tests is to only observe one effect because the modification of the time sequence can go along with a variation of others clock parameters (signal amplitude, laser intensity noise conversion, etc) which also impact the short-term stability. To get rid of this problem, phase noise has been injected into the microwave chain. This ensures that the clock is working in a regime where the Dick effect greatly dominates the total noise budget.

Because of the criticality of the steep decrease of $g(t)$ in the CPT case presented in Section III B (Figs. 4 and 5), tests of detection lengthening and window shaping have been carried out. We first extended the detection duration $\tau_d$, which mainly results in a reduction of the $g(t)$ detection slope, see curves in Fig 8 (top). The starting point of the -40 dB/decade slope of the $g(t)$ Fourier coefficients, seen in Fig. 5, is then shifted toward lower frequencies. This leads to a reduction of sensitivity of the clock to high Fourier frequency noise. Therefore, it lowers the Dick effect when the detection duration increases, as shown in Fig. 8 (bottom). Experimental data and calculated values are in fairly good agreement.

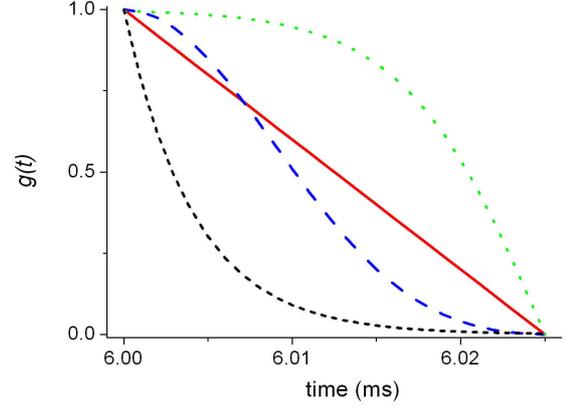

Fig. 9. Sensitivity function, zoomed in on the detection duration, for different theoretical shapes: dots = hyperbolic tangent; solid line = straight line; dashed line = Blackman function;. short-dashed line = exponential function.

The impact of such shapes on the Fourier coefficients of $g(t)$ is depicted in Fig. 10 and a similar behavior for all four shapes is visible below the 60[th] harmonic of the interrogation frequency. For higher harmonics, the hyperbolic tangent and exponential shapes have no added value compared to the linear case. However, the Blackman shape shows a -60 dB/decade slope of great interest because it would strongly weight high frequency noise. Thus the weighting of the detected signal could be interesting, depending on the noise characteristics of the setup. In our case, where the optical PLL bandwidth is around 3 MHz, it would contribute to decrease the Dick effect.

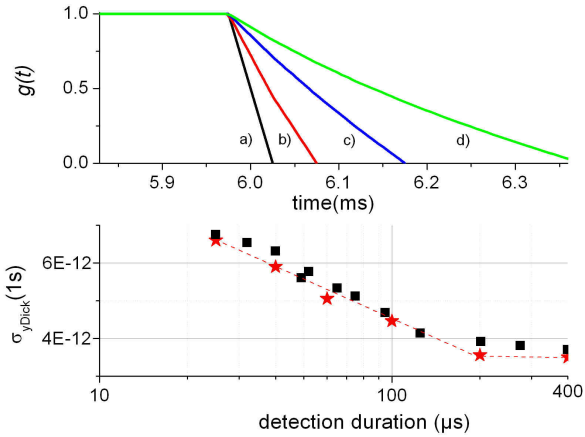

Fig. 8. (top) End part of sensitivity function for different detection durations in CPT case: (a) $\tau_m$ = 25 µs, (b) $\tau_m$ = 75 µs, (c) $\tau_m$ = 175 µs and (d) $\tau_m$ = 375 µs.(bottom): Fractional Allan deviation at 1s as a function of the detection length: squares = experimental values;. stars: calculated contribution of the Dick effect to the fractional Allan deviation at 1s; dashed line = guide tothe eye .

Because the shape of $g(t)$ during detection is important, a theoretical study of how it can be changed, without $\tau_d$ modification, has been undertaken. Various shapes such as an exponential function, Blackman shape [23,24] or a hyperbolic tangent behavior (see Fig 9), can be obtained by applying an appropriate weighting function on the detected signal time series.

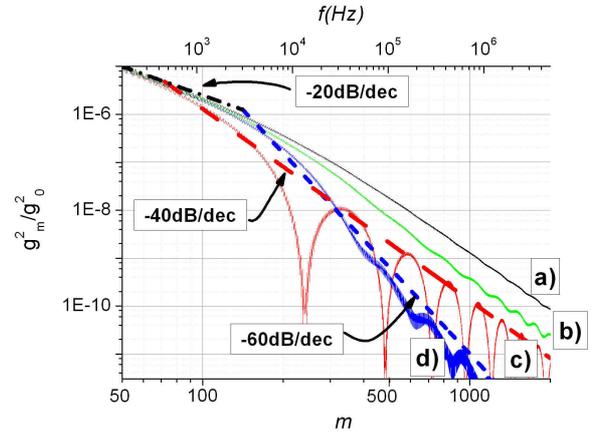

Fig. 10. Fourier coefficient of $g(t)$ for different $g(t)$ shapes during detection: solid line = (a) exponential, (b) hyperbolic tangent, (c) straight line, (d) Blackman; dash-dotted line = asymptote line of low frequency behavior for all shapes at low $m$; short-dashed line: -60 dB/dec asymptote of Blackman shape at high Fourier frequencies; dashed line = -40 dB/dec asymptote of linear shape at high Fourier frequencies.

In a second phase, the duty cycle has been changed by varying the free evolution time $T_R$. The frequency stability at 1 s measured as a function of $T_R$ is shown in Fig. 11 for two values of the injected white phase noise floor, -76 dBrad²/Hz (upper curve, open squares) and -104

dBrad²/Hz (lower curve, solid squares). For small values of $T_R$ the calculated Dick limit of the stability (dots in Fig. 11) is in good agreement with the experimental data. The calculated asymptotic behavior proportional to $T_R^{-1/2}$ is observed (long dashed lines). For higher free evolution time, more than 90% of the atoms have left the CPT state due to relaxation processes, and the clock enters a regime where the signal amplitude becomes very small. Thus, the frequency stability is dominated by the Signal-to-Noise ratio (SNR) here limited by the laser intensity noise. The solid line in Fig. 10 shows the SNR- limited stability.

It is then possible to reduce the Dick limited stability of a CPT clock by optimizing the time sequence which modifies the sensitivity function.

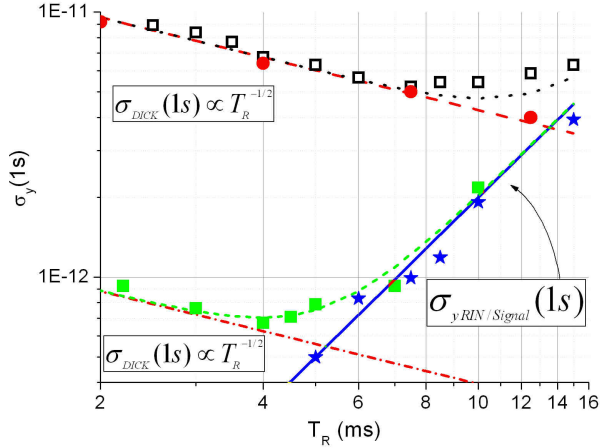

Fig. 11. Frequency stability at 1s as a function of the free evolution time. Two values of the injected white phase noise floor are used: -76 dBrad²/Hz (upper curves, open square) and -104 dBrad²/Hz (lower curves, solid square): squares = experimental data; dots = Dick limit calculated values; short-dashed curves are trend curves; long-dashed lines are a linear fit of the calculated points; stars = calculated SNR stability limit; the solid line is the asymptotic behavior.

## V. Dick effect noise budget

Another way to decrease the Dick effect contribution is to reduce the microwave phase noise. To do so, the noise contribution to the Dick effect of each element of the microwave chain has been evaluated. It turned out that the 100-MHz signal phase noise was the highest contributor because of a high noise level at the clock's most sensitive Fourier frequencies, i.e., the clock interrogation frequency, see dash-dotted curve in Fig. 12. The implementation of a more appropriate 100-MHz quartz oscillator (ULN Rakon, Rackon Ltd., Auckland, New Zealand), allowed its contribution to the Allan standard deviation to be decreased from $6.4\times10^{-13}\ \tau^{-1/2}$ to $1.2\times10^{-13}\ \tau^{-1/2}$.

The phase noise added by the multiplication chain remained unchanged and is characterized by a 1/f slope starting at -80 dBrad²/Hz at 1 Hz and a white phase noise floor at -110 dBrad²/Hz (lowest curve in Fig. 7 (top)). The optical phase locked- loop (OPLL) noise starts to be dominant around 10 kHz, and its phase noise goes up very fast, as $f^{\,4}$. Its contribution to the Dick effect is now the highest of the microwave generation system, evaluated at the $2.1\times10^{-13}$ level. This dominance in the noise budget is well illustrated in Fig. 12, where the measured microwave PSD of the whole chain (square) matches the asymptotic behavior of the OPLL PSD (solid line) for the main part of the Fourier frequencies range.

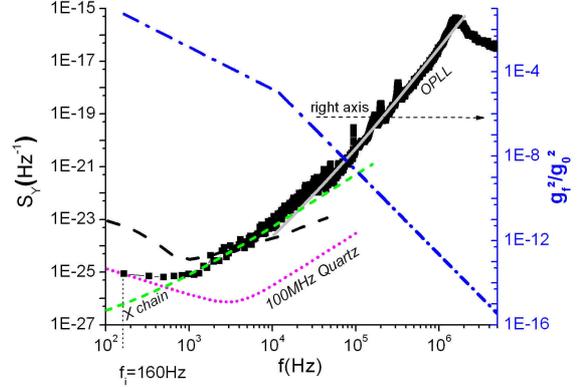

Fig. 12. Left axis: PSD of the relative frequency fluctuations of the free running probe frequency at Fourier frequency $m/T_c$. Dashed line = contribution of the previous 100-MHz LO; dotted line = contribution of new ULN 100-MHz oscillator; short-dashed line = noise contribution of the multiplication chain; solid line = noise contribution of the OPLL; square = total PSD measured at every harmonic of the clock interrogation frequency when the ULN 100-MHz oscillator is used. Right axis, dash-dotted line = Fourier coefficients of $g(t)$. $f_i$ is the interrogation frequency.

The different contributions to the total microwave phase noise are summarized in Table I. The total Dick limited stability is estimated at the level of $2.7\times10^{-13}\ \tau^{-1/2}$. Further improvement requires first a decrease of the noise of the OPLL.

TABLE I. Dick Effect Contributions.

| Device | Range of dominant effect | $S_y$ asymptotic behavior in the frequency range of column 2 | Dick effect contribution |
|---|---|---|---|
| 100MHz | 166Hz to 600Hz | $1.4\times10^{-23}f^{-1} + 3\times10^{-34}f^{\,2}$ | $1.2\times10^{-13}$ |
| X chain | 600Hz to 30kHz | $3\times10^{-29}f + 4\times10^{-32}f^{\,2}$ | $1\times10^{-13}$ |
| OPLL | 30kHz to 5MHz | $2\times10^{-36}f^{\,3} + 4\times10^{-41}f^{\,4}$ | $2.1\times10^{-13}$ |
| Total | 166 Hz to 5MHz | $Sy_{100MHz} + Sy_X + Sy_{OPLL}$ | $2.7\times10^{-13}$ |

To compare our CPT setup with other pulsed microwave clocks having a different phase noise sensitivity, Table II presents the Dick effect calculations for the three cases developed in Section III. Because the pulsed two-level case does not use an OPLL and has a different duty cycle than the CPT case, the noise of the LO is considered to be $S_{y\ 100MHz} + S_{y\ Xchain}$, at the harmonics of its $f_i$, i.e., 215 Hz. The considerable limitation of the Dick effect in the $g(t)$ square window shape highlights the criticality of $g(t)$ slope steepness, wich results in high sensitivity to high-frequency noise. Comparing the pulsed CPT and the pulsed two-level scheme, one sees clearly that the main difference arises from the sensitivity to the multiplication chain and OPLL noise. However, Section IV B has presented different ways for improvement, for example the Blackman shape, which could reduce the impact of those microwave chain elements on the Dick effect, see column 4 of Table II.

TABLE II. Dick Effect Calculation for Some $g(t)$ Functions.

| Device | Square window | Pulsed two-level | Pulsed CPT | Pulsed CPT, Blackman detection |
|---|---|---|---|---|
| 100MHz | 1.67 | 0.9 | 1.2 | 1.2 |
| Xchain | 17.5 | 0.45 | 1 | 1.2 |
| OPLL | 875 | - | 2.1 | 1.1 |
| Total | 875 | 1.0 | 2.7 | 2.0 |

## VI. FREQUENCY STABILITY

With the previous quartz oscillator the measured clock frequency stability was $7\times10^{-13}$ at 1 s, limited by the Dick effect estimated to $6.8\times10^{-13}$ at 1 s, see Fig. 13.

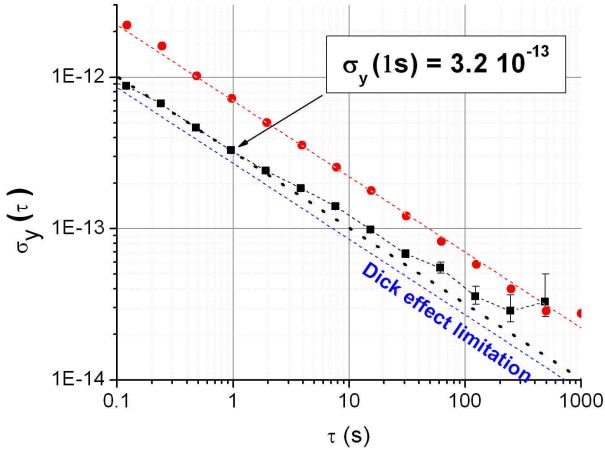

Fig. 13. Short-term frequency stability: dots = measured previous stability; short-dashed line = computed Dick limit at $6.8\times10^{-13}\ \tau^{-1/2}$; squares = measured stability after the 100-MHz quartz oscillator change; dashed line = asymptote of measured value with a $3.2\times10^{-13}\ \tau^{-1/2}$ slope; short-dashed line: Dick effect contribution $2.7\times10^{-13}\ \tau^{-1/2}$.

Thanks to the LO change, the measured stability has been improved to $3.2\times10^{-13}$ at 1 second, a little above the estimated new Dick effect limitation. The Dick effect is still the main limitation of the stability, but the contribution of the signal amplitude noise is no longer negligible. The working parameters are $T_R = 4$ ms, $\tau_m = 25$ µs, without modification of the detection window. Because the Dick effect now only partially limits the clock stability, a full study of the impact of the free evolution time and the detection window on the other parameters like the signal amplitude and the laser intensity noise conversion is compulsory before choosing new working conditions.

## VII. CONCLUSION

We have presented an investigation of the Dick effect for a pulsed CPT vapor cell clock. An original shape of the sensitivity function $g(t)$ has been numerically computed. A simple expression of $g(t)$ is given, providing an easy-to-use tool to estimate the Dick effect on other pulsed CPT systems. Because of the specific shape of $g(t)$, the sensitivity of a pulsed CPT clock to LO noise has been shown to be higher than for a two-level system with a comparable duty cycle. The sensitivity function has been experimentally measured; and it is in very good agreement with the computed one. The validity of the Dick limited frequency stability calculated from $g(t)$ has been verified for different levels of noise floor in the microwave signal. Measured and calculated stabilities are in very good agreement. Thanks to a time sequence optimization, in which the detection duration or the free evolution time are changed, the Dick effect can be reduced. However, in a regime where the Dick effect is no longer the single dominant noise, the impact of such an optimization on the other clock parameters must be studied before its implementation. Finally, this study allowed us to choose an adapted quartz oscillator for our microwave generation system, and it led to a great improvement of the short term stability at the level of $3.2\times10^{-13}$ at 1 second. This value is still partially limited by the Dick effect ($2.7\times10^{-13}$ at 1s) because of the optical-phase-locked-loop noise. Next efforts will focus on understanding the noise-limiting parameters of our PLL. Another solution could be to replace the phase-locked lasers by a single modulated laser. In this case the orthogonally polarized beams can be replaced by a push-pull optical pumping scheme [25-27]. State-of-the-art ultralow-noise quartz oscillators allow a frequency stability lower than $10^{-13}$ at 1 s averaging time for our timing sequence. Without degradation by the PLL and the frequency multiplication chain, this would allow the present Signal-to-Noise limit to be obtained before aiming the shot-noise limit at the $10^{-14}$ level.

## APPENDIX

We detail here the equations used in the numerical computation of the sensitivity function. We consider the two clock levels 1 and 2 of the Cs ground state, separated by the 9.2 GHz hyperfine splitting, and a single excited level 3 coupled to 1 and 2 by optical transitions. The Cs atoms are irradiated by a bichromatic laser field of angular frequencies $\omega_1$, $\omega_2$ close to the atomic resonance frequencies $\omega_{31}$, $\omega_{32}$, respectively, with phases $\varphi_1$, $\varphi_2$. The fast oscillating terms in the Bloch equations are removed by writing the matrix elements in the rotating frame [20]:

$$\rho_{31} = \tilde{\rho}_{31} e^{-i(\omega_1 t + \varphi_1)},$$
$$\rho_{32} = \tilde{\rho}_{32} e^{-i(\omega_2 t + \varphi_2)}, \quad\quad (A1)$$

$$\rho_{21} = \tilde{\rho}_{21} e^{-i((\omega_1-\omega_2)t+\varphi_1-\varphi_2)}.$$

$\rho_{31}$, $\rho_{32}$, $\rho_{21}$ ($\tilde{\rho}_{31}$, $\tilde{\rho}_{32}$, $\tilde{\rho}_{21}$) are the optical and hyperfine coherences, respectively, in the laboratory (rotating) frame. The motion equations of the density matrix elements are written in the rotating wave approximation as:

$$\dot{\rho}_{11} = \Omega_{31}\,\mathrm{Im}(\tilde{\rho}_{31}) + \Gamma_{31}\rho_{33} - r(\rho_{11}-\rho_{22}),$$
$$\dot{\rho}_{22} = \Omega_{32}\,\mathrm{Im}(\tilde{\rho}_{32}) + \Gamma_{32}\rho_{33} - r(\rho_{22}-\rho_{11}),$$
$$\dot{\rho}_{33} = \Omega_{31}\,\mathrm{Im}(\tilde{\rho}_{13}) + \Omega_{32}\,\mathrm{Im}(\tilde{\rho}_{23}) - \Gamma\rho_{33}, \quad (A2)$$
$$\dot{\tilde{\rho}}_{31} = i\frac{\Omega_{31}}{2}(\rho_{33}-\rho_{11}) - i\frac{\Omega_{32}}{2}\tilde{\rho}_{21} + [i\Delta_1-\gamma_{31}]\tilde{\rho}_{31},$$
$$\dot{\tilde{\rho}}_{32} = i\frac{\Omega_{32}}{2}(\rho_{33}-\rho_{22}) - i\frac{\Omega_{31}}{2}\tilde{\rho}_{12} + [i\Delta_2-\gamma_{32}]\tilde{\rho}_{32},$$
$$\dot{\tilde{\rho}}_{21} = -i\frac{\Omega_{32}}{2}\tilde{\rho}_{31} + i\frac{\Omega_{31}}{2}\tilde{\rho}_{23} - [i(\Delta_2-\Delta_1)+\gamma_c]\tilde{\rho}_{21}.$$

$\tilde{\rho}_{ij} = \tilde{\rho}_{ji}^*$. $\rho_{11}$, $\rho_{22}$, $\rho_{33}$ are the populations of the levels 1, 2 and 3, respectively. $\Omega_{31}$, and $\Omega_{32}$ are the Rabi frequencies characterizing the coupling of the 3-1 and 3-2 levels by the field. $\Gamma$ is the relaxation rate of the population of the level 3, $\Gamma_{31}$ and $\Gamma_{32}$ are the relaxation rates of 3 towards 1 and 2, respectively, here $\Gamma = \Gamma_{31}+\Gamma_{32}$, and $\Gamma_{31} = \Gamma_{32}$. $r$ is the decay rate of the difference of population between the levels 1 and 2. $\gamma_{31}$ and $\gamma_{32}$ are the relaxation rates of the optical coherences $\rho_{31}$, $\rho_{32}$. $\gamma_c$ is the relaxation rate of the hyperfine coherence $\rho_{21}$. $\Delta_1$ ($\Delta_2$) is the detuning of the laser defined as: $\Delta_1 = \omega_1-\omega_{31}$ ($\Delta_2 = \omega_2-\omega_{32}$).

In order to compute $\delta S(t, \Delta\varphi)$, (A2) is numerically integrated over the first pulse duration, the free evolution time, and the second pulse duration. A phase step $\Delta\varphi$ between the phases of the two laser fields is introduced at a variable time $t$, the variables $\tilde{\rho}_{31}, \tilde{\rho}_{32}, \tilde{\rho}_{21}$ are modified according to (A1). The detected signal $S$ is the transmitted power though the Cs cell proportional to $-\Omega_{31}\,\mathrm{Im}(\tilde{\rho}_{31}) - \Omega_{32}\,\mathrm{Im}(\tilde{\rho}_{32})$. The proportionality coefficient is irrelevant since the $g_m$ values are normalized to $g_0$ see (1). The numerical values of the parameters used in the computation are: $\gamma_{31} = \gamma_{32} = \Gamma_{31} = \Gamma_{32} = \Gamma/2 = 1.5\times10^9$ s$^{-1}$, $\gamma_c = 0$, $\Omega_{31} = \Omega_{32} = 2.9\times10^6$ rad.s$^{-1}$.

ACKNOWLEDGMENTS

We thank L. Volodimer and J. Pinto for the technical assistance and the realization of various electronic devices. We thank U. Eismann for careful reading of the manuscript. We are grateful to F. Pereira and R. Boudot for helpful discussions and suggestions. We also thank N. Dimarcq, J. Lodewyck and A. Landragin for valuable discussions.